\documentclass[aps,prb,twocolumn,showpacs,showkeys,groupedaddress,floatfix,amsmath,amssymb]{revtex4}
\usepackage{graphicx}
\usepackage{dcolumn}
\usepackage{bm}

\begin{document}


\title{Condition of the occurrence of phase slip centers in
superconducting nanowires under applied current or voltage}

\author{S. Michotte}
\author{S. M$\acute{a}$t$\acute{e}$fi-Tempfli}
\author{L. Piraux}
\affiliation{Unit$\acute{e}$ de Physico-Chimie et de Physique des Mat$\acute{e}$riaux
(PCPM), Universit$\acute{e}$ catholique de Louvain (UCL), Place Croix du Sud
1, B-1348 Louvain-la-Neuve, Belgium}

\author{D. Y. Vodolazov}
\author{F. M. Peeters}
\email{peeters@uia.ua.ac.be} \affiliation{Departement Natuurkunde,
Universiteit Antwerpen (Campus Drie Eiken), Universiteitsplein 1,
B-2610 Antwerpen, Belgium}

\date{\today}

\begin{abstract}
Experimental results on the phase slip process in superconducting
lead nanowires are presented under two different experimental
conditions: constant applied current or constant voltage. Based on
these experiments we established a simple model which gives us the
condition of the appearance of phase slip centers in a
quasi-one-dimensional wire. It turns out that the competition
between two relaxations times (relaxation time of the absolute
value of the order parameter $\tau_{|\psi|}$ and relaxation time
of the phase of the order parameter in the phase slip center
$\tau_{\phi}$) governs the phase slip process. Phase slip
phenomena, as periodic oscillations in time of the order
parameter, is possible only if the gradient of the phase grows
faster than the value of the order parameter in the phase slip
center, or equivalently if $\tau_{\phi}<\tau_{|\psi|}$.
\end{abstract}


\maketitle

\section{Introduction}

Since the discovery of superconductivity it was expected that if
the superconductor was subjected to a constant electric field
superconductivity will inevitably be destroyed. The reason is that
the superconducting electrons will be accelerated by the electric
field and will reach a velocity above its critical velocity.
However if the sample is short enough or if the electric field
exists only in a small part of the sample such that the path over
which the Cooper pairs are accelerated are sufficiently short an
electric field can exist in the sample in the presence of
superconductivity. Another example is the presence of an electric
field in the superconducting sample which is attached to a normal
metal. In this geometry the injected current from the normal metal
will be converted into a superconducting current on a distance of
about the charge imbalance distance ($\Lambda_Q$) and on that
scale an electric field will exist in the sample \cite{Tinkham1}.
In this case the electric field is compensated by the gradient of
the chemical potential $\mu_s$ of superconducting electrons and it
does not lead to an acceleration of the superconducting condensate
(see for example the book of Schmidt \cite{Schmidt}).

But there is another mechanism which allows superconductivity to
survive in the presence of an electric field deep in the
superconducting sample of arbitrary length. This is the {\it phase
slip} mechanism. Initially this phenomena was used in order to
estimate the relaxation time of superconducting current in a
superconducting wire \cite{Langer}. If the order parameter
vanishes in one point of the wire the phase of the superconducting
order parameter exhibits a jump of $2\pi$ at that point
\cite{Langer} and as a result the momentum $p \simeq \nabla \phi$
decreases by $2\pi/L$. So even if the electric field accelerates
the superconducting electrons it does not lead to a destruction of
supercondictivity because the momentum is able to relax through
the phase slip mechanism.

This simple idea was understood already a long time ago.
Therefore, it is very surprising that there are practically no
experimental nor theoretical works studying in detail what will
happen if a voltage (i.e. electric field) is applied to the
superconductor. In previous works mainly the situation with
applied current (i.e. the I=const regime) was studied. In the
latter case the phase slip process was studied theoretically in
detail (see for example the review of Ivlev and Kopnin
\cite{Ivlev} and the books of Tinkham \cite{Tinkham1} and Tidecks
\cite{Tidecks}). On the basis of a numerical solution of the
extended time-dependent Ginzburg-Landau equations it was found
that the phase slip (PS) phenomena exists in some region of
currents. The lowest critical current at which a PS solution first
appears in the system may be smaller than the depairing
Ginzurg-Landau current density and as a result it leads to
hysteresis in the I-V characteristics in the I=const regime
\cite{Ivlev,Tidecks}.

The I=const regime was also studied experimentally in a number of
works \cite{Ivlev,Tidecks} starting from the paper of Meyer and
Minnigerode \cite{Meyer1,Meyer2}. The most characteristic effect
of the phase slip mechanism is the appearance of a stair like
structure in the current-voltage characteristics. In Ref.
\cite{Skocpol} a simple phenomenological model was built in order
to quantitatively describe this feature. It was proposed that
every step in the I-V characteristic is connected with the
appearance of a new phase sip center (PSC) that increases the
resistivity of the sample by a finite value. In this model, it was
conjectured that in a region of the sample with size of about the
coherence length $\xi$ fast oscillations of the order parameter
occurs in time which produces normal quasi-particles. Because of
the finite time needed to convert normal electrons into
superconducting electrons \cite{Tinkham1,Schmidt} there is a
region of size of about $\Lambda_Q$ near the phase slip center
where the electric field and the normal current are different from
zero. It means that the chemical potential of superconducting
$\mu_s$ and normal electrons $\mu_n$ are different near PSC and
their difference is proportional to the charge imbalance Q in a
given point of the superconductor \cite{Tinkham1,Skocpol}. Because
usually $\Lambda_Q \gg \xi$ it is possible to neglect the
$\xi$-sized region with oscillating $|\psi|$ and consider that
phenomena as a time-independent process. Such a model gives a
finite voltage, which is connected with every phase slip center,
which is equal to $V_{PSC}=2\Lambda_Q\rho_n(I-\beta I_{c})/S$ with
$\rho_n$ the normal resistivity, S the cross-section of the
current carrying region, $I_{c}$ is the critical current and
$\beta<1$ is a phenomenological parameter. In the experiment of
Dolan and Jackel \cite{Dolan} the distribution of $\mu_s$ and
$\mu_n$ near a PSC were measured which fully supported the idea of
Ref. \cite{Skocpol} that a difference exist between $\mu_s$ and
$\mu_n$ near the phase slip center.

Despite numerous theoretical and experimental works the physical
conditions under which phase slip phenomena can exist is still not
clear. In a review \cite{Ivlev} on this subject it was claimed
that phase slip phenomena are connected with the presence of a
limiting cycle in the system which leads to such type of
oscillations. But it was not explained how and why it leads to the
phase slip phenomena.

Based on our previous work \cite{Vodolazov2} on the time-dependent
Ginzburg-Landau equations (the investigated systems were
superconducting rings in the presence of an external magnetic
field) we know that systems which are governed by such equations
exhibit two relaxation times. One is the relaxation time of the
phase of the order parameter $\tau_{\phi}$ and the other is the
relaxation time of the absolute value of the order parameter
$\tau_{|\psi|}$. In Ref. \cite{Vodolazov2} it was established that
phase slip processes can occur in such systems when roughly
speaking $\tau_{\phi}<\tau_{|\psi|}$. In the present paper we
discuss this question in the context of superconducting wires in
the presence of an applied current or voltage.

To our knowledge there exists only a single theoretical work in
which a superconducting wire in the presence of an applied voltage
was studied \cite{Malomed}. The authors used the simple
time-dependent Ginzburg-Landau equations and found that the
behavior of the system is very complicated and strongly depends on
the length of the wire and the applied voltage. Neither a detailed
analysis nor any physical interpretation of their results was
presented. In a recent letter \cite{Vodolazov1} we presented our
preliminary theoretical and experimental results on the dynamics
of the superconducting condensate in wires under an applied
voltage. It turned out that in this case the I-V characteristics
exhibit a S-shape which was explained by the appearance of phase
slip centers in the wire and their rearrangement in time. In the
present paper we will present more details and extend our previous
work to the situation in which defects are present, and we
investigate the effects of boundary conditions and an applied
magnetic field.

The paper is organized as following. In Sec. II we present our
theoretical results and give the conditions for the existence of
phase slip centers when current (Sec. IIa) or voltage (Sec. IIb)
is applied to the superconducting wire. In Sec. III we show our
experimental results and in Sec. IV we compare theory and
experiment.

\section{Theory}

We study the current-voltage characteristics of
quasi-one-dimensional superconductors using the generalized
time-dependent Ginzburg-Landau (TDGL) equation. The latter was
first written down in the work of Ref. \cite{Kramer1}
\begin{eqnarray}
\frac{u}{\sqrt{1+\gamma^2|\psi|^2}} \left(\frac {\partial
}{\partial t} +i\varphi
+\frac{\gamma^2}{2}\frac{\partial|\psi|^2}{\partial t}
\right)\psi= \nonumber \\
=(\nabla - {\rm i} {\bf A})^2 \psi +(1-|\psi|^2)\psi.
\end{eqnarray}
In comparison with the simple time-dependent Ginzburg-Landau
equation where $\gamma=0$ it allows us to describe a wider current
region (with a proper choice of the parameters $u$ and $\gamma$)
where a superconducting resistive state exists and gives us a
wider temperature region in which Eq. (1) is applicable
\cite{Kramer1,Kramer2,Watts-Tobin}. The inelastic collision time
$\tau_E$ for electron-phonon scattering is taking into account in
the above equation through the temperature dependent parameter
$\gamma=2\tau_E\Delta_0(T)/\hbar$
($\Delta_0=4k_BT_cu^{1/2}(1-T/T_c)^{1/2}/\pi$ is the equilibrium
value of the order parameter).

Eq. (1) should be supplemented with the equation for the
electrostatic potential
\begin{eqnarray} \Delta \varphi & = & 
{\rm div}\left({\rm Im}(\psi^*(\nabla-{\rm i}{\bf A})\psi)\right),
\end{eqnarray}
which is nothing else than the condition for the conservation of
the total current in the wire, i.e. ${\rm div} {\bf j}=0$. In Eqs.
(1,2) all the physical quantities (order parameter
$\psi=|\psi|e^{i\phi}$, vector potential ${\bf A}$ and
electrostatical potential $\varphi$) are measured in dimensionless
units:  the vector potential ${\bf A}$ and momentum of
superconducting condensate ${\bf p}=\nabla \phi -{\bf A}$ is
scaled by the unit $\Phi_0/(2\pi\xi)$ (where $\Phi_0$ is the
quantum of magnetic flux), the order parameter is in units of
$\Delta_0$ and the coordinates are in units of the coherence
length $\xi(T)$. In these units the magnetic field is scaled by
$H_{c2}$ and the current density by
$j_0=c\Phi_0/8\pi^2\Lambda^2\xi$. Time is scaled in units of the
Ginzburg-Landau relaxation time $\tau_{GL}=4\pi\sigma_n
\lambda^2/c^2=2T\hbar/\pi\Delta_0^2$, the electrostatic potential
($\varphi$), is in units of $\varphi_0=c\Phi_0/8 \pi^2 \xi \lambda
\sigma_n=\hbar/2e\tau_{GL}$ ($\sigma_n $ is the normal-state
conductivity). In our calculations we mainly made use of the
bridge geometry boundary conditions: $|\psi(-L/2)|=|\psi(L/2)|=1$,
$\varphi(-L/2)=0$, $\varphi(L/2)=V$ and
$\psi(L/2,t+dt)=\psi(L/2,t)e^{-i\varphi(L/2)dt}$. We chose these
boundary conditions because at low temperatures the normal current
is converted to a superconducting one due to the Andreev
reflection on a distance of about $\xi_0\simeq 0.18\hbar
v_F/k_BT_c$ near the $S-N$ boundary \cite{Tinkham1,Hsiang}. It
means that practically there is no injection of quasi-particles
from the normal material to the superconductor and hence we can
neglect the effect of charge imbalance near the S-N boundary. The
bridge geometry boundary conditions models this situation. The
parameter $u$ is about $5.79$ according to Ref. \cite{Kramer1}. We
also put $A=0$ in Eq. (1,2) because we considered the
one-dimensional model in which the effect of the self-induced
magnetic field is negligible and we assume that no external
magnetic field is applied.

\subsection{Constant current regime}

Lets consider first the more simple case when a constant external
current is applied to the sample. In such a case it was
theoretically found \cite{Kramer1,Kramer2} that the system
exhibits hysteretic behavior. If one starts from the
superconducting state and increases the current the
superconducting state switches to the resistive superconducting or
normal state at the upper critical current density $j_{c2}$ which,
in a defectless sample, is equal to the Ginzburg-Landau depairing
current density $j_{GL}=\sqrt{4/27}j_0$. Starting from the
resistive state and decreasing the current it is possible to keep
the sample in the resistive state even for currents up to
$j_{c1}<j_{c2}$ (which we call the low critical current). For
$j_{c1}<j<j_{c2}$ such a state is realized as a periodic
oscillation of the order parameter in time at one point of the
superconductor \cite{Kramer1,Kramer2}. When the order parameter
reaches zero in this point a phase slip of $2\pi$ occurs. This is
the reason why such a state is now called a phase slip state and
this point a phase slip center (PSC). Using results obtained in
our earlier work \cite{Vodolazov2} we claim that the value of
$j_{c1}$ depends on the ratio between the two characteristic times
in the sample: the phase relaxation time of the order parameter
$\tau_{\phi}$ and the relaxation time of the absolute value of the
order parameter $\tau_{\psi}$ in the region (with size of about
$\xi$) where the oscillations of the order parameter occurs.

Having written Eq. (1) for the dynamics of the phase and the
absolute value of the order parameter in a quasi-one dimensional
wire of length L (-L/2$<$s$<$L/2)
\begin{subequations}
\begin{eqnarray}
u\sqrt{1+\gamma^2|\psi|^2}\frac {\partial |\psi|}{\partial t} &  
 = &  \frac{\partial^2 |\psi|}{\partial s^2}+|\psi|(1-|\psi|^2-p^2),
\qquad
\\
\frac {\partial \phi}{\partial t} & = &
\varphi-\frac{\sqrt{1+\gamma^2|\psi|^2}}{u|\psi|^2}\frac{\partial
j_n}{\partial s},
\qquad 
\end{eqnarray}
\end{subequations}
it is easy to estimate both these times. Indeed, from Eq. (3a) it
directly follows that
\begin{equation} 
\tau_{|\psi|}\sim u\sqrt{1+\gamma^2|\psi|^2}\simeq u\gamma \quad
({\rm for} \gamma \gg 1).
\end{equation}
To determine $\tau_{\phi}$ we need to know how fast the phase
changes over the region (with size of about $\xi$) at which the
order parameter oscillates. Our numerical calculation shows that
the time-average of the derivative is $\partial \langle \phi
\rangle /\partial t=C \theta(s)$ (with C(s)=const and $\theta(s)$
the theta function). Because the time-averaged electrostatic
potential $\varphi$ changes over a distance $\Lambda_Q \gg 1$ for
$\gamma \gg 1$ (where
$\Lambda_Q^2=\sqrt{1+\gamma^2|\psi|^2}/u|\psi|^2 \simeq \gamma/u$
is the decay length of the normal current density and the charge
imbalance $Q$) and that the order parameter is about unity already
at $s=\pm\xi$, we can estimate this constant as $C\simeq
\Lambda_Q^2\partial j_n/\partial s\simeq j/\Lambda_Q$.
Consequently we find that
\begin{equation} 
\tau_{\phi}\simeq \frac{1}{\Lambda_Q j} .
\end{equation}
The product $\Lambda_Q j$ is roughly the voltage drop $V_{PSC}$
'produced' by the phase slip center \cite{Skocpol}. Thus the time
change of the phase of the order parameter in the phase slip
center is inversely proportional to the voltage drop over the
whole structure (and $C=V_{PSC}/2$). This is a consequence of the
fact that the time-averaged electro-chemical potential of the
superconducting electrons $\mu_s^e=\partial \langle \phi \rangle
/\partial t$ does not change in space and its jump near the phase
slip center is equal to the voltage drop of this structure in the
sample. In terms of the language used by Schmid and G. Sch\"on
\cite{Schmid}, $\tau_{|\psi|}$ is the 'longitudinal' time and
$\tau_{\phi}$ decreases with increasing 'transverse' time because
$\tau_{\phi}\sim 1/\Lambda_Q \sim 1/\sqrt{\tau_Q}$. In Ref.
\cite{Vodolazov2} it was found that the phase slip events occur
periodically in time when $\tau_{\phi} \lesssim \tau_{|\psi|}$.
This gives us an estimation for the lower critical current
\begin{equation} 
j_{c1}\simeq \frac{1}{\tau_{|\psi|}\Lambda_Q}.
\end{equation}

By varying the parameter $\gamma$ we can change both
$\tau_{|\psi|}$ and $\Lambda_Q$ and hence we can vary $j_{c1}$. In
Fig. 1 we show the voltage in the sample, the normal current
density and the absolute value of the order parameter in the phase
slip center averaged over the time as a function of the external
current for two different values of $\gamma$. In our simulations
we started from the superconducting state with $j<j_{GL}$. At
$j>j_{GL}$ the system jumps instantaneous to the resistive state
with finite voltage \cite{self1}. Than we decrease the external
current and at $j<j_{c1}$ the system transits to the purely
superconducting state with $V=0$. We should emphasize that at $j
\to j_{c1}$ the voltage jumps by a finite value (this result is
qualitatively different to the results of Ref.
\cite{Kramer1,Watts-Tobin} where the authors found that $V\to 0$
at $j\to j_{c1}$.). It means that there is a non-infinite maximal
oscillation period for the order parameter in the phase slip
center. We believe that the finite voltage jump $\Delta V$ or
finite period of oscillations is directly connected with the
threshold condition $\tau_{\phi}\simeq \tau_{|\psi|}$ for
activation of regular phase slip processes in the constant current
regime and it means that $\Delta V \sim 1/\tau_{|\psi|}$.

Before going further we should stress here that the above
condition for the existence of a phase slip process
$\tau_{\phi}\lesssim \tau_{|\psi|}$ is a rather rough estimate.
Indeed, Eqs. 3(a,b) are a coupled system of equations, and besides
$\tau_{|\psi|}$, it explicitly (see Eq. (4)) depends on the value
of $|\psi|$. However, the above condition allows us to explain the
general qualitative properties of the phase slip processes
(including the existence of $\Delta V$, $j_{c1}$ and its
dependence on $\gamma$ and $u$) and predict new features which
will be discussed below.

\begin{figure}[hbtp]
\includegraphics[width=0.48\textwidth]{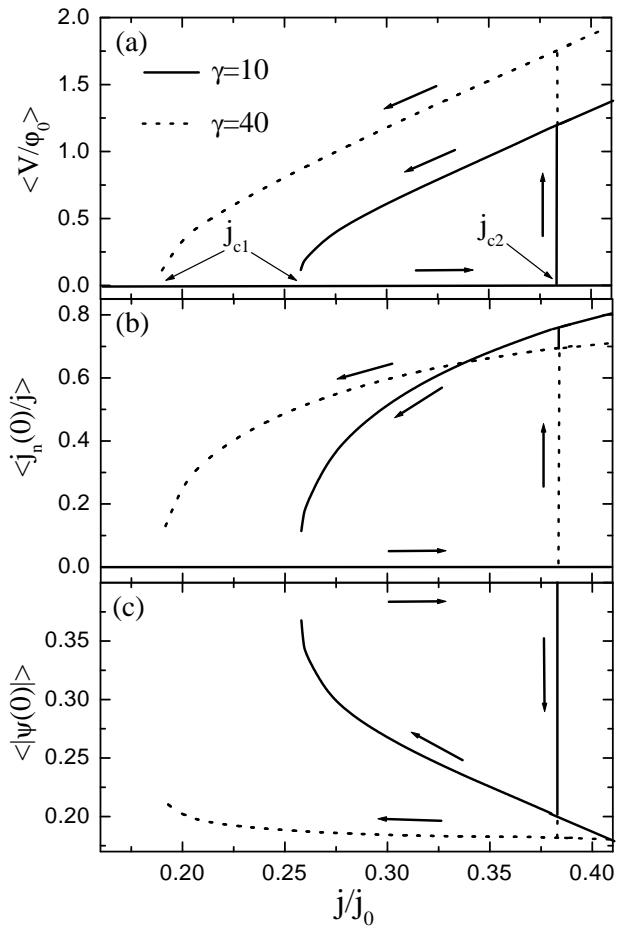}
\caption{The dependence of the time averaged voltage (a) on the
external current for a wire containing only one phase slip center.
In figures (b,c) the dependencies of the normal current density
(b) and the order parameter (c) in the phase slip center are
shown. Length of the wire is $40\xi$. Solid curves correspond to
$\gamma=10$($\Lambda_Q\simeq 2.3\xi$), and the dotted curves are
for $\gamma=40$ ($\Lambda_Q\simeq 4.1\xi$).}
\end{figure}

Far enough from $j_{c1}$ the dependence of $V(j)$ on the current
is close to linear. When the current increases, $\tau_{\phi}$
decreases (see Eq. (5)) and hence the time $\tau_{PSC}$ between
two phase slips also decreases and the order parameter has less
time for recovering at the phase slip center. That is the reason
why the time averaged voltage increases (as $\langle
V\rangle=2\pi/\tau_{PSC}$), the averaged order parameter
$\langle|\psi(0)|\rangle$ decreases and the fraction of the normal
current $\langle j_n(0)\rangle/j $ increases with increasing
external current. It is interesting to note that from the
phenomenological Skocpol-Beasley-Tinkham (SBT) \cite{Skocpol}
model it follows that $\langle j_n(0) \rangle/j=1-\beta
{j_{c1}}/j$ which qualitatively resembles the dependence shown in
Fig. 1(b).

Let us now discuss the effect of defects on the I-V
characteristic. This question was considered previously in Ref.
\cite{Kramer3} for two different models of defects: local
variation of the critical temperature and a local variation of the
mean free path. We will repeat these calculations partially and
interpret it in terms of a competition between $\tau_{\phi}$ and
$\tau_{|\psi|}$. In addition to the first type of defect we will
also study the effect of the variation of the cross-section of the
wire.

The first type of defect is the inclusion of a region in the
superconductor which suppresses $T_c$ and the order parameter
becomes lower than the equilibrium value $\Delta_0$ even in the
absence of any external current. This is modelled \cite{Kramer3}
by introducing the term $\rho(s)\psi$ in the right hand side (RHS)
of Eq. (1). In the present calculation we choose
$\rho(s)=-\rho_0\theta(0.5-|s|)$. The larger $\rho_0$ the more the
order parameter is suppressed in the center of the wire. Firstly,
such a defect leads to a decrease of the upper critical current
$j_{c2}$. Secondly, it decreases the lower critical current
$j_{c1}$. Indeed, when we introduce a defect, the RHS of Eq. (3a)
decreases at the spatial position where the order parameter
oscillations and hence $|\psi|$ needs more time to change in that
spatial position. Thus this type of defect leads to an increase of
the relaxation time of the order parameter in the region of the
defect (as an indirect prove of this we found a decreasing $\Delta
V$ with increasing 'strength' of the defect). If the size of the
defect is smaller than $\Lambda_Q$ we can neglect its effect on
the relaxation time of the phase of the order parameter. Finally,
we can conclude that the 'stronger' the defect the smaller the
value of the lower critical current $j_{c1}$.

\begin{figure}[hbtp]
\includegraphics[width=0.48\textwidth]{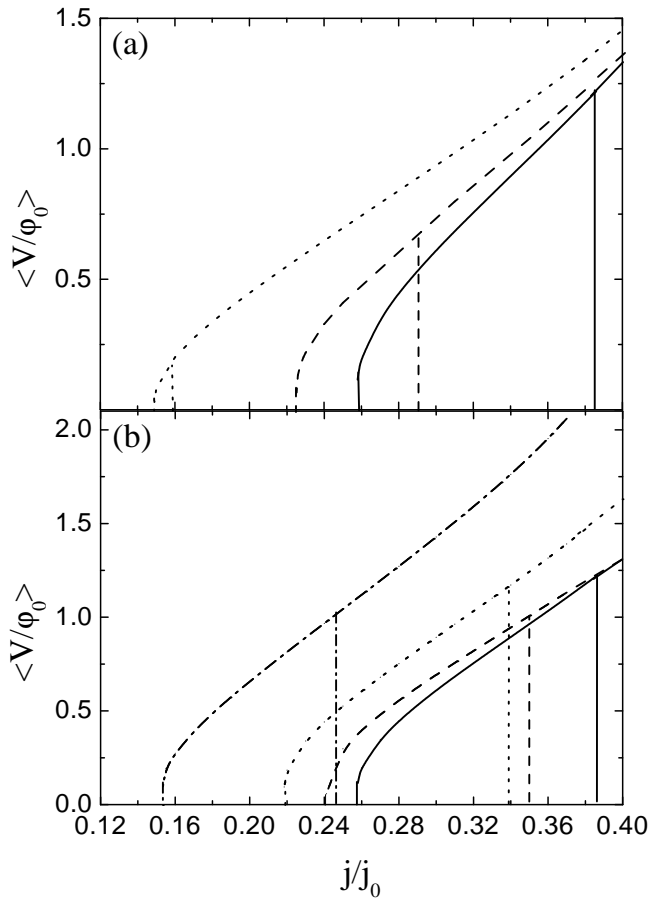}
\caption{The dependence of the time averaged voltage on the
external current for a wire containing a single defect. Figure (a)
corresponds to a wire with a local suppression of the $T_c$
(dotted curve for $\rho_0=-2$, dashed curve for $\rho=0$ and solid
curve for a wire without defect). Figure (b) corresponds to a
local variation of the cross-section of the wire which we modelled
as $D(s)=1-\beta e^{-s^2/l_d^2}$. Solid curve in this figure is
for a wire without a defect, dashed curve for a wire with defect
parameters $\beta=0.5$ and $l_d=0.5$, dotted curve with
$\beta=0.2$ and $l_d=2$, dash-dotted curve with $\beta=0.5$ and
$l_d=2$. The length of the wire is $L=40\xi$ and the parameter
$\gamma=10$ ($\Lambda_Q\simeq 2.3\xi$).}
\end{figure}

In Fig. 2(a) we present our numerical results for two different
values of $\rho_0$. With increasing 'strength' of the defect the
upper and lower critical currents decrease and start to merge. As
a consequence the hysteresis in the I-V characteristics will
disappear when the defect is sufficiently strong.

The second type of defect is one for which we have a local
decrease of the cross-section of the wire. In this case we expect
that $\tau_{|\psi|}$ will not be influenced. The situation with
$\tau_{\phi}$ is more complicated because a large part of the
superconductor with size of about $\Lambda_{Q}$ participates in
the formation of this time. Two limiting cases can be
distinguished. If the defect is much smaller than $\Lambda_Q$ we
can neglect its effect on $\tau_{\phi}$ and hence we will have the
same lower critical current as for the case of an ideal wire. In
the opposite case of a defect with size much larger than
$\Lambda_{Q}$ we only have to take into account the increased
current density in the part of the wire where we have a smaller
cross-section and apply Eq. (5). In this case the current $j_{c1}$
is decreased by a factor $D_{av}/D_{l}$, where $D_{av}$ is the
average cross-section and $D_l<D_{av}$ is the local reduced
cross-section.

Another important question is the value of the upper critical
current $j_{c2}$. If the size of the defect is larger than $\xi$
then the proximity effect from adjacent parts near the defect will
be small and $j_{c2}$ will decrease by a factor $D_{av}/D_l$ (in
the opposite case the current $j_{c2}$ is almost defect
independent). Therefore, for a defect with length $\xi< l \ll
\Lambda_Q$ the I-V characteristic may be reversible with a proper
choice of the parameters (like in the case of a local suppression
of $T_c$). In Fig. 2(b) we show the results of our numerical
calculations and we obtain a qualitative agreement with the above
physical arguments. When the length of the defect is smaller than
$\Lambda_Q$ then the lowest critical current density practically
does not change. In the opposite case $j_{c1}$ decreases by a
factor of $D_{av}/D_l$. The upper critical current density changes
considerably only if the length of the defect exceeds $\xi$.
Unfortunately, we are not able to consider the case for which the
length of the defect is simultaneously much smaller than
$\Lambda_Q$ and much larger than $\xi$ because of computational
restrictions. For example, when we increase $\gamma$ by a factor
of two the time of calculations increases also by a factor of two
but the ratio $\Lambda_Q/\xi$ increases roughly only by a factor
of $\sqrt{2}$ for $\gamma \gg 1$. The reason is that when we
increase $\gamma$ we need to take a smaller time step (see Eq.
(1)) which increases the computation time.
\begin{figure}[hbtp]
\includegraphics[width=0.48\textwidth]{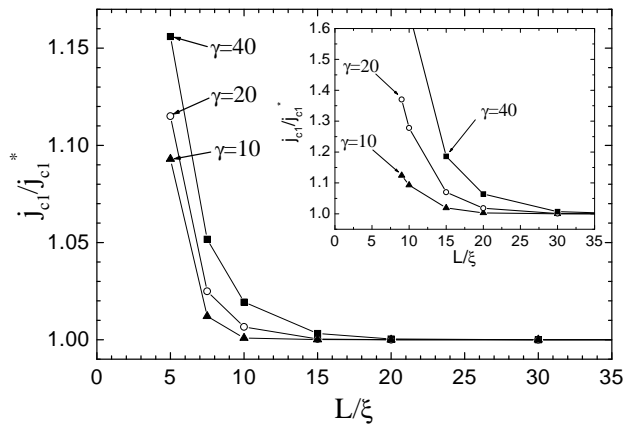}
\caption{Dependence of the low critical current $j_{c1}$ on the
length of the superconducting wire for different values of
$\gamma$. The current $j_{c1}$ is normalized to its value
($j_{c1}^*$) at lengths $L>>\Lambda_Q$. With increasing $\gamma$
the decay length of the normal current increases and hence
$j_{c1}$ starts to depend on $L$ for longer wires. In the inset,
the dependence of $j_{c1}$ on the length of the superconducting
wire for S-N boundary conditions is shown.}
\end{figure}

When the wire has a length less than $\Lambda_Q$ then this will
affect the distribution of the normal current density in the wire
and hence the time $\tau_{\phi}$ and the lower critical current
$j_{c1}$. Indeed from our bridge boundary conditions it follows
that $\partial j_n/\partial s(\pm L/2)=0$ (see Eq. 3(b)). Then
taking into account that $\partial \langle\phi\rangle/\partial
t=V\theta(s)/2$ we can easily obtain, in the limit $\Lambda_Q \gg
\xi$, that
\begin{equation} 
\langle j_n(0)\rangle=\frac{V(j)}{2\Lambda_Q}\frac{1}{{\rm
tanh}(L/2\Lambda_Q)},
\end{equation}
where the current $\langle j_n(0)\rangle$ is the average normal
current in the phase slip center.

The voltage jump $\Delta V \simeq 2\pi/\tau_{|\psi|}$ at $j_{c1}$
should not change with varying $L$ (at least for $L \gg \xi$ when
proximity effect does not have an effect on $\tau_{|\psi|}$). From
Eq. (7) it follows then that the normal current density increases
with increasing length. But $\langle j_n(0)\rangle$ should always
be smaller than the full current $j$. It implies that the critical
current density $j_{c1}$ will increase with increasing length of
the wire as $j_{c1}\sim1/ {\rm tanh(L/2\Lambda_Q})$ in order to
keep the ratio $\langle j_n(0)\rangle/j$ constant. Fig. 3
illustrates the dependence of $j_{c1}$ on the length of the wire
which we obtained on the basis of a numerical solution of Eqs.
(1,2). Unfortunately, in our calculations we are not able to use
very large values of $\gamma$ and the maximal value of
$\Lambda_{Q}$ was $4.1\xi$ for $\gamma=40$. But nevertheless, we
found that $j_{c1}$ increases with decreasing wire length. We
should add here that we also found that $\Delta V$ also decreases
a little bit. We connect it with a small change in $\tau_{|\psi|}$
due to the small length of the wire and hence the increased effect
of the boundaries.

The more pronounced effect of the finite length of the wire on the
value of $j_{c1}$ is found for the case that we used N-S boundary
conditions: $\psi(\pm L/2)=0$ and $\partial \varphi/\partial s
(\pm L/2)=-j$. These boundary conditions are approximately valid
for samples at temperatures close to $T_c$. It is easy to show
that in this case
\begin{equation} 
\langle j_n(0)\rangle=\frac{V(j)}{2\Lambda_Q} \frac{{\rm
tanh}(L/2\Lambda_Q)}{1-1/\alpha\cdot {\rm cosh}(L/2\Lambda_Q)},
\end{equation}
where $\alpha=\langle j_n(0)\rangle/j<1$ (see Fig. 1(b)). In this
case $j_{c1}$ also increases with decreasing wire length (see
inset in Fig. 3). But besides there is a finite length $L_0$ for
which $\langle j_n(0)\rangle \to \infty$. It implies that the
phase slip process is not possible in wires with length $L<L_0$.
In such wires the system goes from the superconducting state
directly to the normal state.

Not only the finite length of the sample is able to change the
value of $j_{c1}$. If we apply a magnetic field parallel to the
length of the wire it will suppress the order parameter in the
sample. Our analysis shows that if the diameter of the wire $d$ is
less than $2\xi$ and if we can neglect screening effects
($\lambda>\xi$) then the distribution of the order parameter will
be uniform along the cross-section of the wire. The order
parameter depends on H as
$$
|\psi|^2=1-(H/H_c)^2
$$
with $H_c\simeq 2.9\Phi_0/\pi \xi d$. This behavior is very
similar to the behavior of a thin plate in a parallel magnetic
field \cite{Ginzburg,Douglass} or a thin and narrow ring in a
perpendicular magnetic field \cite{Vodolazov3}. In all cases the
transition to the normal state is of second order and the
vorticity in the wire will be equal to zero due to the small
cross-section of the sample.

Because the order parameter practically does not depend on the
radial coordinate of the wire we can use the one-dimensional model
in order to study the response of the system on the applied
current. In order to take into account the suppression of $|\psi|$
by magnetic field ($H$) we add to the RHS of Eq. (1) the term
$-(H/H_c)^2\psi$. In some respect it is similar to the way we
introduced the first type of defect in our wire. We can expect
that $\tau_{|\psi|}$ will increase with increasing $H$ (because
the 'strength' of the 'defect' increases). But because the
magnetic field suppresses the order parameter everywhere in the
sample it also leads to an increase of $\Lambda_Q$, because in the
TDGL model $\Lambda_Q \sim 1/\sqrt{|\psi|}$. It is clear that both
these processes should decrease $j_{c1}$. The strongest mechanism
is connected with the change in $\tau_{|\psi|}$. Indeed it is easy
to estimate that $\tau_{|\psi|}\simeq 1/(1-(H/H_c)^2)$ and
$\tau_{\phi} \simeq (1-(H/H_c)^2)^{1/4}$. Even if we take into
account that the parameter $\gamma$ may decrease with increasing H
(because in non-zero magnetic field there is another pair-breaking
mechanism and instead of $\tau_E$ we should
use\cite{Schmid,Kadin1} $\tau_E/\sqrt{1+2\tau_E\tau_s}$ with
$\tau_s=\Delta(T=0,H=0)/\hbar (H/H_c(T=0))^2$ ) it does not lead
to an increase of $j_{c1}$ with an increase of $H$ because
$\tau_{|\psi|}$ changes faster than $\tau_{\phi}$ even in this
case.

In the above model it is easy to show that
$j_{c2}(H)=\sqrt{4/27}(1-(H/H_c)^2)^{1.5}$ for the case of an
uniform wire. In Fig. 4 we present the results of our numerical
calculations. We found that both $j_{c1}$ and $j_{c2}$ decreases
with increasing magnetic field and at some $H^*$ they practically
merge. Unfortunately, it is quite difficult to find an analytical
expression for the dependence of $j_{c1}(H)$ like we had for
$j_{c2}$. The reason is the complicated behavior of the dynamics
of $\psi$ in the phase slip center.

The variation $\Lambda_Q$ with increasing $H$ was obtained
experimentally in Ref. \cite{Kadin1}. To compare with the theory
the authors of Ref. \cite{Kadin1} used the expressions found in
the work of Schmid and Sch\"on \cite{Schmid} which are valid in
the limit $T\to T_c$. It is interesting that in Ref. \cite{Kadin1}
it was found that expressions of Schmid and Sch\"on are
quantitatively valid even far from $T_c$. From this observation we
may hope that the present theoretical results are also valid over
a wider temperature range than only near $T_c$.

\begin{figure}[hbtp]
\includegraphics[width=0.48\textwidth]{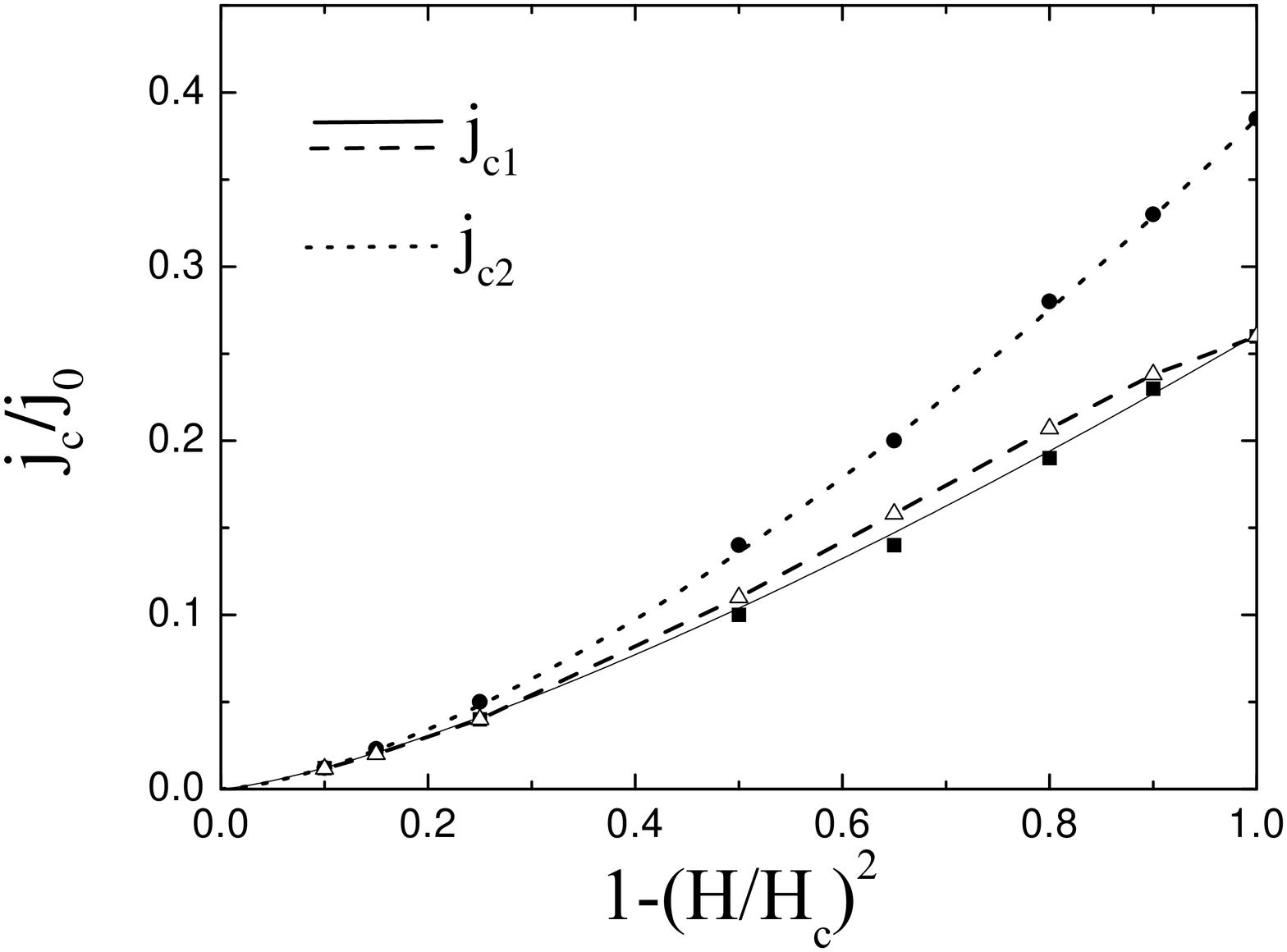}
\caption{Dependence of the lower critical current $j_{c1}$
(squares and solid curve) and upper critical current $j_{c2}$
(dots and dotted curve) on the applied magnetic filed. Dotted
curve is equation $\sqrt{4/27}(1-(H/H_c)^2)^{1.5}$. The solid
curve is the fitted expression $0.26\cdot(1-(H/H_c)^2)^{1.36}$. We
also plotted the dependence $j_{c1}(H)$ (open triangles and dashed
curve) where the effect of $H$ on $\gamma$ was taken into
consideration. We used the simple expression
$\gamma_{eff}=\gamma_0/\sqrt{1+\gamma_0(H/H_c)^2}$ which even
overestimates the effect of $H$ (i.e. we took
$\gamma_0=\gamma(H=0)$).}
\end{figure}

The main conclusion which follows from the change of $j_{c1}$ with
decreasing wire length and/or applying magnetic field is that there
exist a critical length $L^*$ (or critical field $H^*$)
below(above) which the current $j_{c1}$ becomes equal to $j_{c2}$.
It implies that for wires with lengths $L<L^*$ and/or fields
$H>H^*$ there will be no jump in the voltage in the
current-voltage characteristic and the I-V curve will be
reversible. It will also result in the absence of a S-behavior in
the V=const regime (see section below).

\subsection{Constant voltage regime}

In our earlier work we found that the I-V in the $V=const$ regime
exhibits an S-like behavior \cite{Vodolazov1} and furthermore for
low voltages there is an oscillatory dependence of the current on
the applied voltage (see Fig. 5). As was shown in Ref.
\cite{Vodolazov1} these properties are connected with the
existence of two critical currents $j_{c1}$ and $j_{c2}$. Here, we
will discuss the characteristic voltages $V_1$, $V_2$ (see Fig. 5)
and their dependence on the length of the sample.
\begin{figure}[hbtp]
\includegraphics[width=0.48\textwidth]{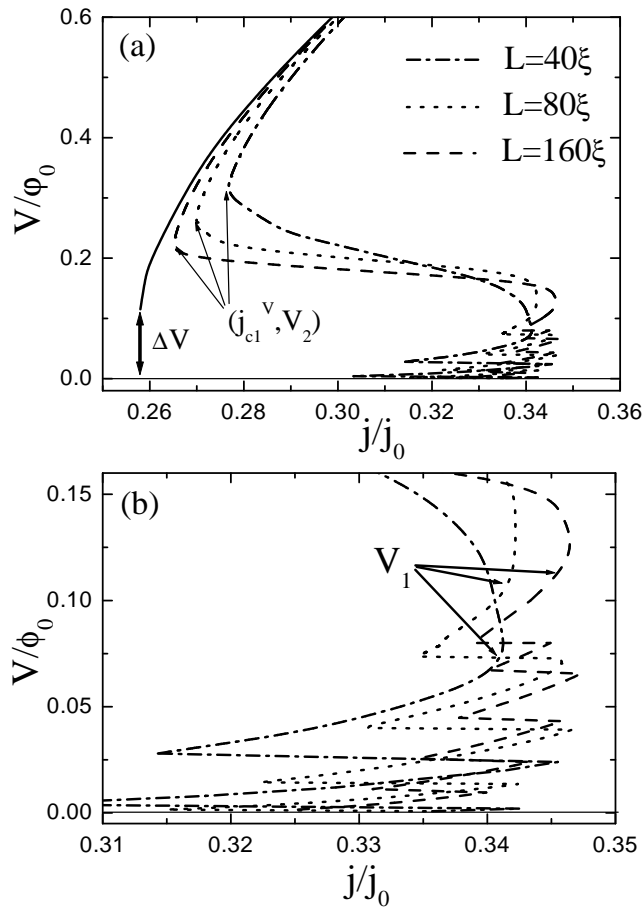}
\caption{Theoretical current-voltage characteristics of wires of
different lengths for $\gamma=10$. Solid curve corresponds to the
I=const regime and is practically universal for the considered
lengths. Dot-dashed curve ($L=40 \xi$), dotted curve ($L=80\xi$)
and dashed curve ($L=160 \xi$) correspond to the $V=const$ regime.
(b) is an enlargement of low voltage region in (a).}
\end{figure}

If we apply a voltage $V$ to the wire of length $L$ then in the
sample an electric field $E=V/L$ exist which will accelerate the
superconducting electrons. When the current density approaches
$j_{c2}$ phase slip centers will spontaneously appear in the
sample. As a result the momentum of the superconducting electrons
(and hence the current) will decrease by $2\pi/L$ after each phase
slip event. When the current density decreases below $j_{c1}$ the
phase slip process will no longer be active and the applied
electric field will be able to accelerate the superconducting
condensate. This process leads to periodic oscillations in time of
the current in the sample.

We can divide the period of oscillations (at least at voltages
$V<V_1$) in two parts. First, the longest part is the one during
which the condensate is accelerated by the electric field till the
moment reaches $j \simeq j_{c2}$. The second part we call the {\it
transition period} $T_{tr}$. The transition period also consists
of two parts: the time needed for the phase slip processes (which
is proportional to the number of phase slip events and hence the
length of the wire) and the time $T_0$ for the decay of the order
parameter from $\sqrt{2/3}$ (when $j\simeq j_{c2}$) till the first
phase slip event and for the recovering of $|\psi|$ back to
$\sqrt{2/3}$ from zero after the last phase slip event (see Fig.
6). The minimal number of phase slip events which occur during the
transition time is determined by the internal parameters of the
superconductor ($j_{c1}$) and its length \cite{Vodolazov1} which
is given by
\begin{equation}
{\rm N_{min}=Nint\left((p_c-p_{c1})(L/2\pi+1)\right)},
\end{equation}
where $p_{c1}$ is the smallest real root of the equation
$j_{c1}=p_{c1}(1-p_{c1}^2)$ and $\rm{Nint(x)}$ returns the nearest
integer value.

At $V<V_1$ the  time-averaged current $\langle j \rangle$
increases (with oscillations) and in the range $V_1<V<V_2$ it
decreases with increasing voltage. The lowest minimal current in
the latter region is $j_{c1}^V$ which depends on the length of the
system (see Fig. 5 and Ref. \cite{Vodolazov1}). In Fig. 7 we
present the dependence of the above voltages on the length of the
wire. The explanation for their different behavior is the
following. At a voltage $V_1$ the period of the oscillation
$T=2\pi N/V_1$ (N is the number of phase slip events during the
transition time) of the current becomes of order $2T_{tr}$. The
time $T_0$ does not depend on the length and the voltage at $V\sim
V_1$. So, we can estimate $V_1$ as
\begin{equation} 
V_1\simeq \frac{2\pi N}{2T_{tr}}= \frac{\pi N}{T_0+\langle
\tau_{PSC} \rangle N}=\frac{\pi\alpha L}{T_0+\langle \tau_{PSC}
\rangle\alpha L},
\end{equation}
where $\langle \tau_{PSC} \rangle$ is the average time between two
phase slip events and $\alpha$ is the coefficient which depends on
$p_{c1}$ (see Eq. (9)) and hence on the specific superconductor.
When $T_0 \ll \langle \tau_{PSC} \rangle \alpha L$ the voltage
$V_1$ becomes practically independent of the length. Because the
time $T_0$ depends on the relaxation time of the absolute value of
the order parameter the length of the wire $L_{sat}$ at which
$V_1$ saturates depends on the internal parameters of the
superconductor.

\begin{figure}[hbtp]
\includegraphics[width=0.48\textwidth]{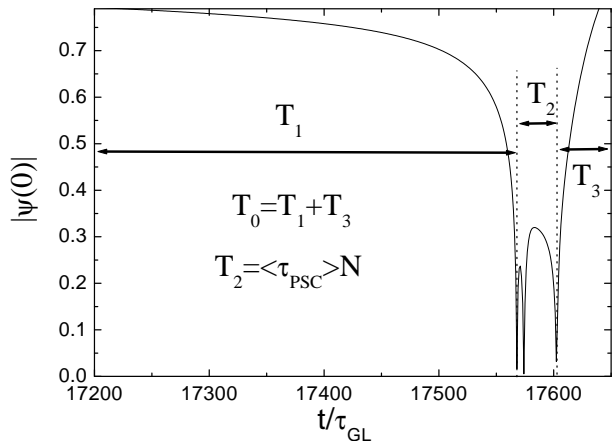}
\caption{The transition time consists of two parts:
$T_{tr}=T_0+T_2$.}
\end{figure}

\begin{figure}[hbtp]
\includegraphics[width=0.48\textwidth]{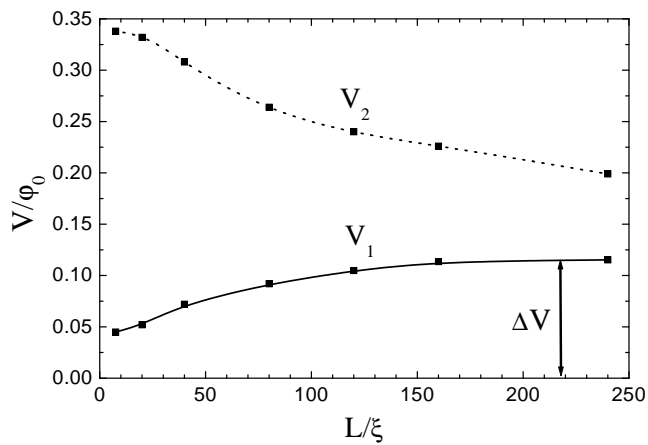}
\caption{Dependencies of the voltages $V_1$ and $V_2$ on the
length of the wire. The results are obtained for $\gamma=10$.}
\end{figure}

The voltage $V_2$ decreases with increasing length of the sample
because the lower critical current $j_{c1}^V$ decreases. It is
interesting to note that with accuracy of finding $V_1$ the
saturated value of the lower critical voltage coincides with the
voltage jump $\Delta V$ at $j=j_{c1}$ in the constant current
regime. Because the minimal value of $V_2$ is equal to $\Delta V$
for an infinitely long wire we may conclude that with increasing
wire length the I-V characteristic in the range of voltages
$(V_1,V_2)$ approaches the horizontal line.

Defects, magnetic field, short length of the sample etc. will
decrease the hysteresis and consequently the currents $j_{c1}$ and
$j_{c2}$ approaches each other. If the difference between them
will be small enough the S-shape of the I-V characteristic at the
constant voltage regime changes to the usual monotonic behavior
(see Fig. 8). This happens because when the voltage approaches
$V_1$, the maximal current during the period of the current
oscillation $T$ will be about $j_{c1}$ (because $j_{max}$ cannot
substantially exceed $j_{c2}$ but $j_{c2}$ is already close to
$j_{c1}$ in this case).

\begin{figure}[hbtp]
\includegraphics[width=0.48\textwidth]{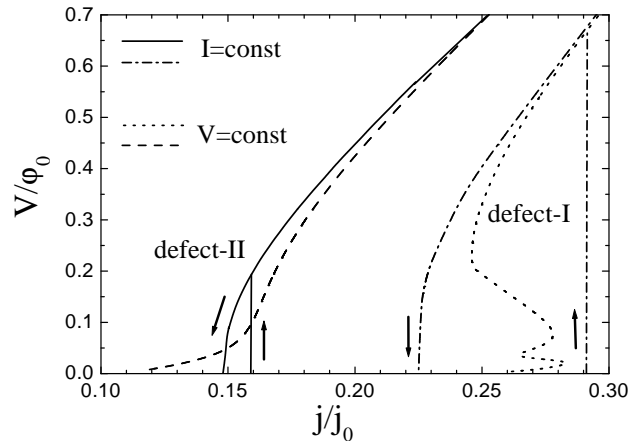}
\caption{Current voltage characteristics of a superconducting wire
with a local suppression of the critical temperature in the center
of the wire. The parameters for the defect and the wire are the
same as in Fig. 2(a). For the I-V characteristic with small
hysteresis in the $I=const$ regime the I-V behavior in the
$V=const$ regime is a single-valued function of the current.}
\end{figure}

We also would like to discuss how the change of the boundary
conditions will change the shape of the I-V characteristic. If we
apply the N-S boundary conditions the I-V curve in the voltage
driven regime also exhibits an S-behavior but without oscillations
in the current at small voltages (see Fig. 9). In this case there
will always be an inevitable voltage drop near the boundaries
connected with the current in the wire through the relation
$V_{NS}\sim j\Lambda_Q$ (because the electric field and the normal
current decays on a scale of the charge imbalance length
$\Lambda_Q$ near the boundaries \cite{Tinkham1}). Near the
boundary the nonzero electric field is compensated by the term
$\Lambda_Q^2 \partial E/\partial s$ (see Eq. 3(b) - and $E=j_n$ in
our units) and the superconducting electrons are not accelerated
by this field. The situation is very similar to the case when we
inject a current in the wire. When the current generated by this
voltage reaches $j_{c1}$ the phase slip process becomes possible
in the system. But like the case of the current driven regime the
superconducting state can be stable(metastable) until the current
reaches $j_{c2}$ at $V\simeq j_{c2}\Lambda_Q$. From our estimates
it follows (see section below) that the fluctuations of the order
parameter are very important in our samples. Therefore, in our
calculations we introduce at some moment of time (at fixed
voltage) a phase slip center in the center of the wire and checked
if it will survive or not. As a result we obtained the I-V
characteristics as presented in Fig. 9. When the voltage is less
than some critical value ($V^*$) the phase slip process decays in
time and the resulting current in the wire is time-independent.
The whole voltage drop occurs near the boundaries. At $V>V^*$ (and
$j>j_{c1}$) the dynamics of the condensate in the wire will be
similar to the case considered above for $V>V_1$.

The reason for this is as follows. The current density in the wire
at small voltages will always be less than $j_{c1}$ and $j_{c2}$
due to the relation $j\sim V_{SN}/\Lambda_Q=V/\Lambda_Q$. When the
current density reaches $j_{c1}$ the phase slip process becomes
possible in the sample. But such a phase slip process leads to a
finite voltage. As a consequence the voltage drop at the
boundaries will sharply decrease when a phase slip center is
created. But it implies that the full current density will also
decrease and consequently will become less than $j_{c1}$. We can
conclude that the phase slip process may survive in such a type of
superconductor only if the applied voltage will be roughly equal
to $\Delta V(j)$ plus the voltage drop near the boundaries
$V_{SN}$ necessary for the creation of a current larger than
$j_{c1}^V$.

\begin{figure}[hbtp]
\includegraphics[width=0.48\textwidth]{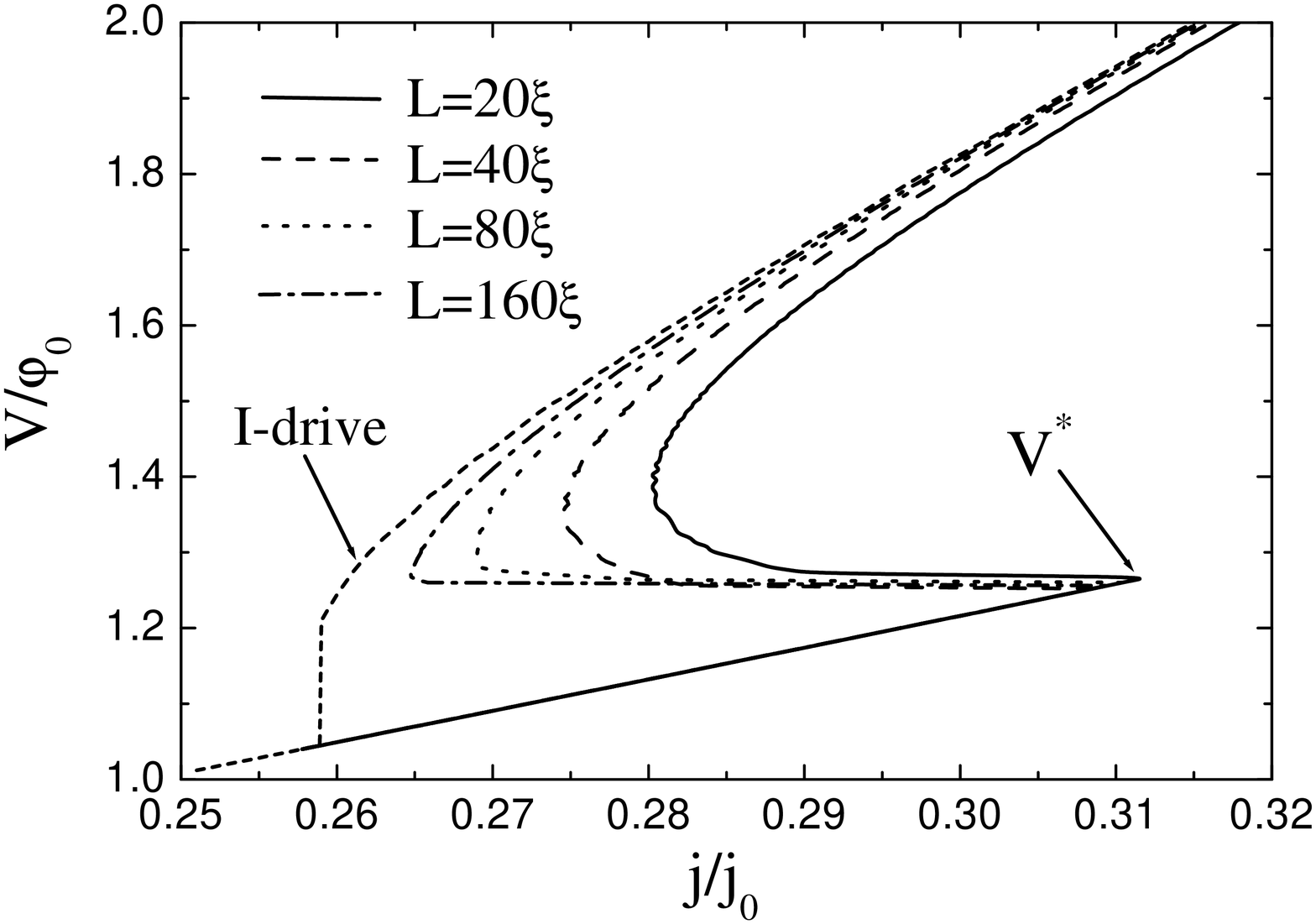}
\caption{Current-voltage characteristics of wires of different
lengths in the case of the N-S boundary conditions.}
\end{figure}

Concluding the theoretical part, we present in Fig. 10 the I-V
characteristics in the I=const and V=const regimes at different
temperatures close to $T_c$. With decreasing temperature the range
of currents, where there is a S-behavior increases and the
voltages $V_1$ and $V_2$ increases. It resembles the experimental
results presented in \cite{Vodolazov1,Michotte} and in the
subsequent Section III.

\begin{figure}[hbtp]
\includegraphics[width=0.48\textwidth]{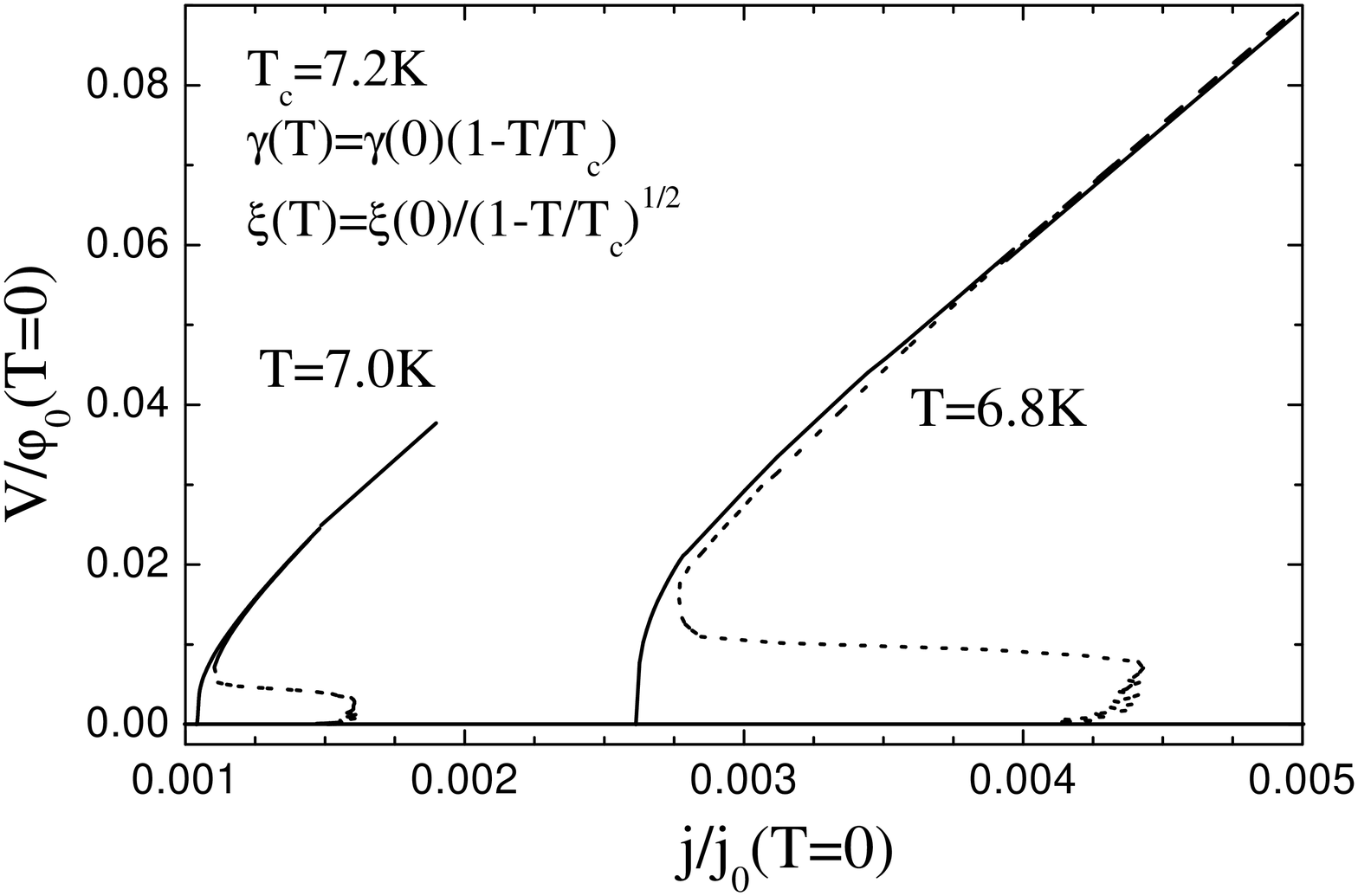}
\caption{Current voltage characteristics of a superconducting wire
of length $600\xi(0)$ at different temperatures close to $T_c$. We
used typical parameters for Pb: $\xi(0)\simeq 40 nm$ and
$\gamma(0)\simeq 100$. This result shows an increase of the
difference $j_{c2}-j_{c1}$ in absolute units with decreasing
temperature.}
\end{figure}

\section{Experiment}

\begin{table*}
\caption{Parameters of the different samples.}
\begin{ruledtabular}
\begin{tabular}{ccccccccc}
& L$(\mu m)$ & d(nm) & R$_n$(7.1K)($\Omega$) &
R$_{res}$(4.3K)($\Omega$) & H$_c$(4.5K)(T) & $\xi_1$(4.5K)(nm) &
$\xi_2$(4.5K)(nm) &
$\rho_n$(7.1K)($\mu \Omega$ $\cdot$ cm) \\

\hline
A & 22 & 40 & 300.9 & 14.9 & 1.271 & 37 & 16 & 1.67  \\
B & 50 & 55 & 210.2 & 21.7 & 0.925 & 38 & 19 & 0.86  \\
C & 50 & 55 & 465.9 & 80.7 & 1.652 & 21 & 14 & 1.75  \\
D & 50 & 70 & 94.6  & 17.6 & 0.591 & 46 & 24 & 0.52  \\

\end{tabular}
\end{ruledtabular}
\end{table*}

Measurements of the I-V characteristics were done on single Pb
nanowires \cite{Vodolazov1,Michotte} with typical diameter about
$50 nm$ and length $22-50 \mu m$ (see Table I). At first we would
like to discuss the rather wide resistive transitions and strong
magneto-resistance effect (see Figs. 11-12) in our samples. We may
claim, taking into account the small diameter of our nanowires,
that the effect of the thermo-activited \cite{Langer} and the
quantum-activated \cite{Giordano,Tinkham2} phase slip phenomena is
very strong in our samples. It is easy to see that with increase
of the diameter of the nanowire the width of the resistive
transition decreases. Our estimations, based on Giordano
expressions, \cite{Giordano} showed that for sample A even at
$T=0$ the number of quantum phase slip events should be about
$\sim 10^5$ per second. That is the reason why we did not observe
in our experiment any hysteresis in the current driven regime (see
below). But this rate is not large enough for destroying the
S-behavior in the voltage driven regime. Indeed, the period of
oscillations of the current for Pb is about $10^{-9}$s at $V\sim
V_1$ and $T=0$. It means that only at temperatures close to $T_c$
the fluctuating PSC's will interrupt the internal temporal
oscillations in the order parameter and ruin the S-behavior. In
other words the system can be in a state with $j>j_{c1}$ only
during a time which is less than the time between two phase slips.
This results in the coincidence of the I-V characteristics both in
the current and the voltage driven regimes at temperatures close
to $T_c$ (see Fig. 3 in Ref. \cite{Michotte}). We would like also
to mention the small difference in the critical temperatures which
implies that all our samples have practically the same value of
the superconducting gap.

\begin{figure}[hbtp]
\includegraphics[width=0.48\textwidth]{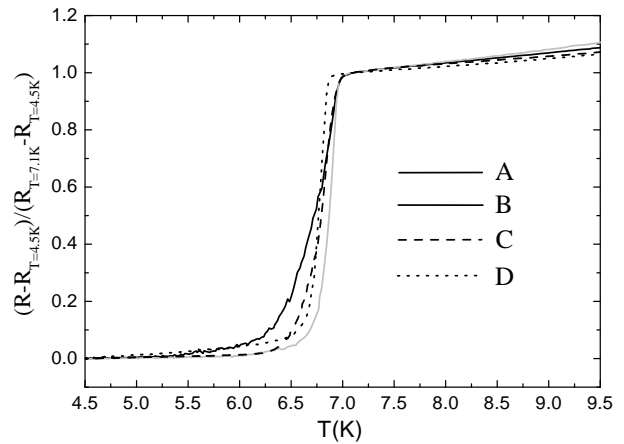}
\caption{Resistive transition of our different samples. Note that
for the narrowest sample (A) the width of the transition is widest
and for the widest sample (D) it is narrowest.}
\end{figure}

As can be seen in Fig. 12, the wider the transition width in
$R(T)$, the smaller the magnetic fields where the resistance
starts to increase with increasing magnetic field.

\begin{figure}[hbtp]
\includegraphics[width=0.48\textwidth]{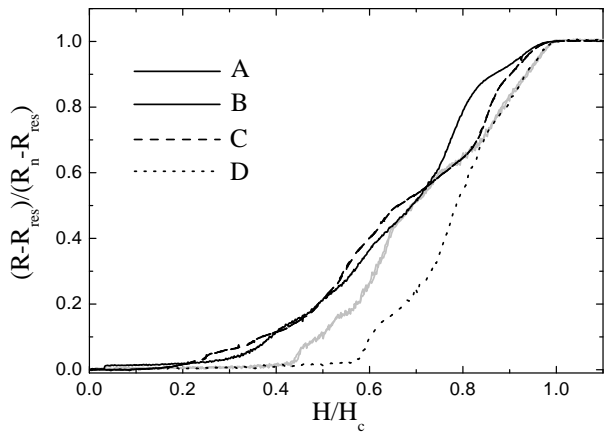}
\caption{Magneto-resistance of our different samples at 4.3 K.}
\end{figure}

Of course, our nanowires are not free from imperfections. Although
our fabrication technique produce samples which are quite regular
(see Ref. \cite{Michotte} for Scanning Electron Microscopy (SEM)
picture), we may distinguish two kinds of imperfections. Firstly,
there can be deviations from an ideal cylindrical shape. Indeed,
although the nanopores of the membrane in which the nanowires are
grown are designed to be as regular as possible, there may exist a
smooth variation in diameter along the length of the nanowire,
which we estimated to be not more than 10 $\%$. However, in
addition to these smooth variations, it may happen that a defect
in the membrane leads to a constriction in the measured nanowire.
In this case, the diameter can be locally reduced quite
importantly and this results in a lower critical current and a
larger critical magnetic field. This is in fact what we observed
in sample C. In addition to such shape imperfections, we note also
that the parameters of our nanowires strongly vary from sample to
sample - see table I (we estimated the coherence length using the
expression $H_c=2.9\Phi_0/(\pi \xi d)$ ($\xi_1$) and
$H_c=\Phi_0/(2\pi\xi^2)$ ($\xi_2$) for the critical field). The
reason for this difference in resistivity is related to the second
kind of imperfections, which is structural disorder that is formed
during the electrodeposition of these nanowires inside the
nanopores. Indeed, as shown in Ref. \cite{Dubois}, these nanowires
are polycrystalline and inevitably contain structural defects such
as dislocations, twins, etc. These two kinds of imperfections
results in differences in the I-V characteristics (Fig. 13). For
example for samples A and B we observed two jumps in the voltage
which we explain by the appearance of two successive phase slip
centers in the nanowire but for sample C we found only one jump in
the voltage. Indeed, due to the presence of a constriction in this
sample, heating may drastically affect the behavior of this
sample, precipitating the return to the normal state.
Unfortunately sample D was broken during the measurements and we
could not measure its I-V characteristic.

\begin{figure}[hbtp]
\includegraphics[width=0.48\textwidth]{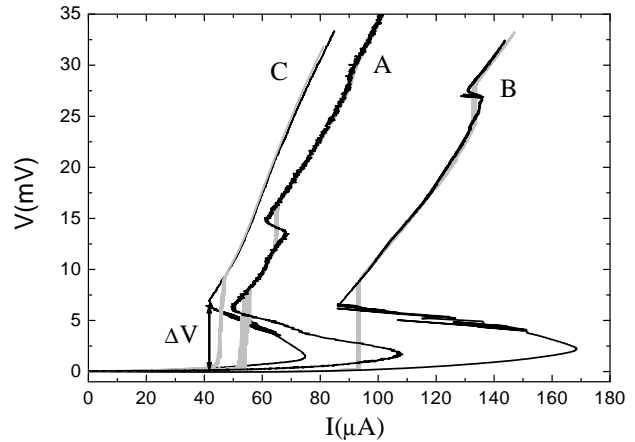}
\caption{Current-voltage characteristics of samples A(T=4.3 K),
B(T=4.37 K) and C(T=4.2K) in the current (grey curves) and voltage
(black curves) driven regimes. The contact voltage was subtracted
from the experimental data.}
\end{figure}

We found that jumps in the voltage $\Delta V$ are practically the
same for all three samples (see Figs. 13-14). It agrees with the
theory presented in Sec. II. Indeed all the three samples have
almost the same $T_c$ and hence the same superconducting gap. It
is naturally also to suppose that the time $\tau_E$ is almost the
same for all our samples. We can expect that the parameters
$\gamma$ and $u$ are the same for all our samples and consequently
the relaxation time $\tau_{|\psi|}$ and the ratio $\Lambda_Q/\xi$
are the same. This automatically leads to the invariance of
$\Delta V$ on $\rho_n$. From the results of Sec. IIb follows that
the voltage $V_2$ decreases with increasing (in $\xi$) nanowire
length (see Fig. 6). For our shortest sample A: L$\simeq 600-1400
\xi$ (see Table I). It means that in our nanowires the voltage
$V_2$ already reaches the minimal value $\Delta V$ and hence $V_2$
is independent of $\rho_n$ for our nanowires. The voltage $V_1$
depends not only on the length of the nanowire but also on the
time change of the order parameter during the transition period
$T_0$ (see Eq. (9)) and hence the voltage $V_2$ may reach $\Delta
V$ at shorter lengths than $V_1$ if the time $T_0$ is large enough
(in Fig. 6 the opposite situation is presented - at first $V_1$
reaches the saturated value).

If the current density is uniformly distributed over the
cross-section of the sample then the oscillations of the order
parameter will be in phase along the cross-section and in this
case the $\Delta V$ will not depend on the size of the sample. It
is a direct consequence of the fact that the time-averaged
chemical potential of the superconducting electrons should be
equal and constant on both sides of the phase slip center or phase
slip line/surface.

Our results show if one uses the Skocpol-Beasley-Tinkham (SBT)
\cite{Skocpol} model for the estimation of $\Lambda_Q$ one should
be very careful. Indeed, those authors replaced the normal current
density in the PS center by the expression $I_n(0)=(I-\beta
I_{c1})$. As a result the derivative $dI_n(0)/dI=1$ becomes
current independent. But in general the time derivative may be
large than unity (see Figs. 1(a,b)). Secondly, if the length of
the sample is comparable with $\Lambda_Q$ we should replace
$2\Lambda_Q$ by $2\Lambda_Q {\rm tanh}(L/2\Lambda_Q)$ (see Eq.
(7)). As a result the differential resistance in samples with
$L\sim \Lambda_Q$ may be larger, in general, than the normal one
\cite{ours2}. This case corresponds to our samples ($R_{dif}\simeq
320$ Ohm, $R_{dif}\simeq 351$ Ohm and $R_{dif}\simeq 456$ Ohm for
samples A,B,C respectively after the first jump in voltage). But
from the SBT model follows that $R_{dif}\leq R_{n}$ (the equality
sign holds for samples with $L\lesssim \Lambda_Q$).

Unfortunately we do not know the actual dependence $I_n(0)(I)$ for
our samples. If we use the values obtained from the SBT model
($\Lambda_Q=12.6 \mu m$, $48.5 \mu m$ and $30.3 \mu m$ for samples
A,B and C, respectively) which is qualitatively understandable for
longer samples, we have a too small number of PSC (compare samples
A and C). And in this case, the question arises why $\Lambda_Q$
changes so much. Probably, the SBT model gives us only the correct
order of magnitude for $\Lambda_Q$ which is only useful as an
estimate.

\begin{figure}[hbtp]
\includegraphics[width=0.48\textwidth]{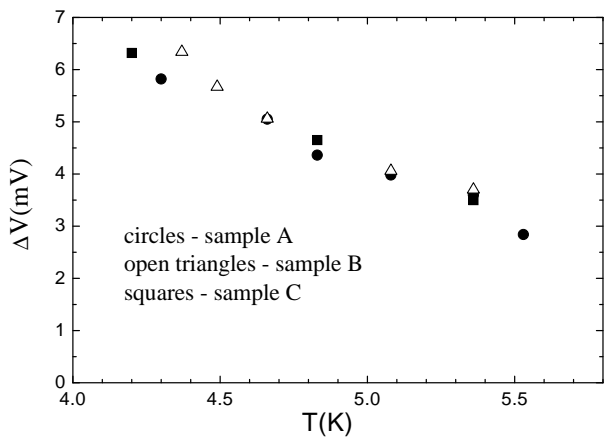}
\caption{Jump in the voltage $\Delta V$ (see Fig. 5(a) and 13) for
samples A-C at different temperatures.}
\end{figure}

Finally, we present our results on the influence of an applied
magnetic field on the I-V characteristics. We limit ourselves to
data for sample A in the current driven regime (see Fig. 15).
These results already support our theoretical predictions of
Section IIb. It is evident that the lower critical current density
decreases with increasing applied magnetic field. From some range
of H-values only one jump in the voltage exists. We explain it by
an increase of $\Lambda_Q$ at relatively large magnetic fields and
hence there is a lack in space for the coexistence of two phase
slip centers in the nanowire. At high magnetic fields the order
parameter is strongly suppressed by H and the effect of quantum
phase-slip fluctuations becomes more pronounced. This is the
reason for a smoothing of the I-V characteristics at high magnetic
fields.

\begin{figure}[hbtp]
\includegraphics[width=0.48\textwidth]{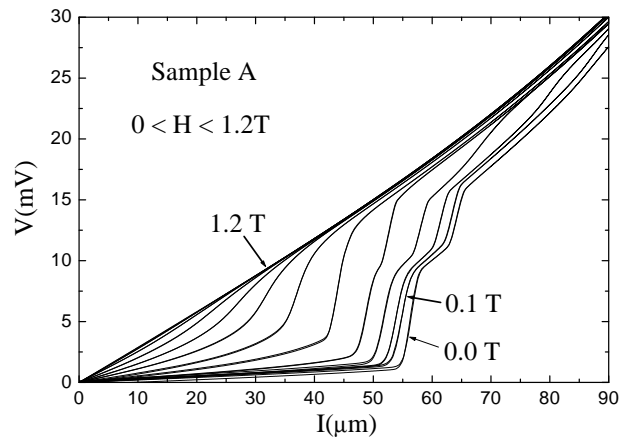}
\caption{Current-voltage characteristics of sample A (T=4.3 K) in
the current driven regime in the presence of a parallel magnetic
field. The magnetic field increases from right to left with step
of $0.1 T$.}
\end{figure}

\section{Discussion}

In conventional superconductor near T$_c$ the N-S boundary
conditions (in the sense mentioned in Sec. IIa) are valid
\cite{Hsiang}. For $T\to 0$ the bridge boundary conditions are
more applicable due to Andreev reflections at the border between
the normal metal and the superconductor \cite{Andreev,Hsiang}. At
intermediate temperatures we expect a mixture of the N-S and
bridge boundary conditions. It means that part of the voltage will
drop at the boundaries and part in the sample. The situation in
our experiment is even more complicated because in our two-point
measurements there is also a voltage drop at the contacts.
Probably, this will not allow us to observe the oscillations of
the current in the voltage driven regime. Another reason is that
our samples are very long compared to $\xi$ (even for our 22 $\mu
m$ sample $L\simeq 1000 \xi$) and consequently the amplitude of
the oscillations is very small. But nevertheless our calculations
in both limiting cases of bridge and N-S boundary conditions
predicts an S-shape of the I-V characteristic.

We found that the type of boundary conditions are not so important
for the process of nucleation of phase slip centers if the length
of the sample is much larger than $\Lambda_Q$. However, the
difference in the process of the conversion of superconducting
electrons to normal ones and vice versa at the N-S boundary at
various temperatures plays a crucial role for the creation of
phase slip centers in shorter nanowires. It turned out that for
similar parameters (nanowire's length, coherence length,
superconducting gap, $\tau_E$) phase slip centers appear in the
superconducting wire at smaller currents for the case of the
bridge geometry boundary conditions.

All our theoretical results are strictly speaking only valid near
$T_c$, which is the temperature region where Eqs. (1,2) are
quantitatively correct. Nevertheless, experimental results
supports our prediction on the competition of two relaxation times
in the creation of a phase slip center even far from $T_c$.
Indeed, an applied parallel magnetic field decreases the lower
critical current (compare Fig. 15 and Fig. 4). Jumps in the
voltage $\Delta V$ turned out to be almost an independent function
of the disorder (of the resistance of the sample) as it follows
from theory. Unfortunately, it is rather difficult to check the
dependence of $j_{c1}$ on the length of the nanowire using our
technique because every preparation of a new sample leads to a
different level of disorder and hence different values for $\xi$
and $\Lambda_Q$.

Therefore, we expect only quantitative differences in the
dependence of $j_{c1}(T,H)$, $j_{c2}(T,H)$, as compared with our
theoretical results. Some qualitative differences (see Ref.
\cite{Baratoff}) in the dynamic of the order parameter at the
phase slip center or the creation of charge imbalance waves
\cite{Kadin2} cannot affect the main properties of our theoretical
results because it does not influence the existence of the two
different critical currents: $j_{c1}$ and $j_{c2}$. It may lead to
quantitative differences in the dependence of $\tau_{\phi}$ and
$\tau_{|\psi|}$ on the microscopic parameters of the
superconductor.

Finally, we would like to discuss other mechanisms which, for our
geometry, may lead to an S-behavior of the I-V in the V=const
regime. The first is heating \cite{Volkov}. In order to explain
the double S-structure for our samples A, B by this mechanism we
have to assume that heat dissipation and heat evacuation have a
very complicated and non trivial dependence on temperature. We do
not know any mechanisms which can lead to such adependence in our
case. Besides we did not any observe hysteresis in the current
driven regime which is an inevitable property of that mechanism if
the I-V characteristic would have a S-shape in the voltage driven
regime. For these reasons we believe that heating is not
responsible for the observed behavior.

Secondly, if the nanowire contains a normal region the I-V
characteristic will exhibit an S-behavior due to multiple Andreev
reflection in the SNS structure \cite{Nicolsky}. We do not have
any indication for this process from our R(T) and R(H)
measurements which shows that our samples are quite homogeneous.
Furthermore the structure would occur at voltages much smaller
than $\Delta/e$. In our case the S-behavior is seen for
$V>\Delta/e$ - see Fig. 13 (for Pb, $\Delta(0)\simeq 1.4 meV$).
Andreev reflection on the N-S boundaries may give rise to a zigzag
shape of the I-V in the V-const regime \cite{Sols} but this effect
is negligible for our samples with $L\gg \xi_0$.

\begin{acknowledgements}

This work was supported by IUAP(P5/1/1), GOA (University of
Antwerp), ESF on "Vortex matter", the Flemish Science Foundation
(FWO-Vl) and the "Communaut\'e Fran\c caise de Belgique" through
the program "Actions de Recherches Concert\'ees". D.V. is
supported by DWTC to promote S $\&$ T collaboration between
Central and Eastern Europe. We thank the POLY lab. at UCL for the
polymer membranes. S. Michotte is a Research Fellow of the FNRS.

\end{acknowledgements}


\begin{references}
\bibitem{self1} For our parameters $L/\Lambda_Q\gg 1$ so that in order to prevent
the appearance of several phase slip centers simultaneously when
$j>j_{c1}$ we put a small defect (of first type with $\rho_0\ll 1$
- see discussion in the text on the effect of defects on the phase
slip process) in the center of the sample in order to provide the
existence of only one PSC in the wire.
\bibitem{ours2} It is interesting to note that if in a short
sample $L\sim \Lambda_Q$ two PS centers may appear and the
differential resistance will practically not change after every
voltage jump and should be approximately equal to the normal one
if $dI_n(0)/dI \simeq 1$. The change in the voltage will probably
only be connected in this case with a change in the coefficient
$\beta$, because in short nanowires the PSC should strongly
interact with each other. If the coefficient $\beta$ does not
change it would be quite difficult to notice the appearance of the
second phase slip center.
\end{references}
\end{document}